\documentclass[preprint,showpacs,preprintnumbers,amsmath,amssymb]{revtex4}
\usepackage{graphicx}
\usepackage{bm}
\usepackage{subfigure}
\begin{document}
\topskip 20mm
% \preprint
% \tightenlines
%\title{Universality in Complexity: An accuracy based Diffusion Model}
%\title{(Search for) Towards a common thread in Complexity: An accuracy based approach}
\title{ Towards a common thread in Complexity: an accuracy-based approach}
\author{Pragya Shukla}
\affiliation{Department of Physics,
Indian Institute of Technology, Kharagpur, India.}
\date{\today}
% Time-stamp: <00/05/24 08:40:44 somen>
% \onecolumn
\widetext

\begin{abstract}
% . .   
   
     The complexity of a system, in general,  makes it difficult to determine 
     some or almost all matrix elements of its operators. The lack of accuracy 
     acts as a source of randomness for the matrix elements which are also 
     subjected to an external potential due to existing system conditions. The 
     fluctuation of accuracy due to varying system-conditions leads to a diffusion 
     of the matrix elements. We show that, for single well potentials, 
     the diffusion can be described by a common 
     mathematical formulation where system information enters through a 
     single parameter. This further  leads to a characterization of 
     physical properties by an infinite range of single parametric universality 
     classes.

%    This further leads to a single-parametric formulation 
%     of the statistical behavior of the physical properties.     

%     Here we consider the cases where the 
%     system conditions can be modeled by  a multiparametric, monotonic external 
%     potential. 

%     A recent study shows that the statistical properties of complex systems 
%     can be described by a common mathematical formulation where system information 
%     enters through a single parameter. This leads to a characterization of 
%     physical properties by an infinite range of single parametric universality 
%     classes. The formulation is based on the random matrix modeling of complex 
%     systems using maximum entropy hypothesis. This paper 
%     presents an alternative approach for the formulation 
%     and generalizes its applicability to a wider range of complex systems. 
   
\end{abstract}

 
%..
%\begin{multicols}{2}

\maketitle

\section {Introduction}
 .

For systems that can be described mathematically, physical information 
can be derived, in principle, from a detailed knowledge of the operators 
that govern their evolution. Physical systems can however 
be complex in nature and it is not always possible to determine the operator 
exactly or, even if they are known, to solve the equations they determine.
This paper aims to model  
the statistical behavior  of those complex systems where a matrix 
representation of the operators is meaningful. 

The complexity may appear in various forms, for example, 
as noise due to many body interactions or external disorder potential, 
as chaos due to scattering of a particle from boundaries 
(e.g. clean quantum dots), 
as coherence patterns emerging out of randomness (see, for example,  
\cite{kwa2} for various definitions of complexity). 
For example, consider the Hamiltonian 
 of a many body system . If the local interactions are complicated in a 
specific part of the system, the evaluation of the corresponding matrix 
elements becomes technically difficult. 
These elements  can then be determined only 
within a certain degree of accuracy and  can best be described by 
a probability density. However the system may also contain parts 
where interactions are simple and the related matrix elements 
can exactly be calculated. The operator then turns out to be a matrix 
with both random and non-random elements; we refer such a matrix 
as a generalized random matrix.  Simlar matrices would also appear for 
systems containing a combination of chaotic as well as ordered 
components. The properties of such system can then be modeled by an 
ensemble of generalized random matrices.

In recent years, due to increasing degree of complexity in systems of 
industrial and technological interests, the mathematical models such 
as random matrix ensembles have become necessary. In fact, a particular 
class of these ensembles, known 
as stationary ensembles \cite{me}, have been successfully applied for 
modeling of the operators for  a wide range of complex 
systems e.g. nuclei, atoms, molecules, disordered and chaotic systems, 
quantum chromodynamics, elastomechanics, electrodynamics (see reviews 
\cite{gu,me,iz, fren2, been, alh, bohi, ve, wig, weid} and references therein 
for details), mathematical areas such as Riemann zeta function, enumeration 
problems in geometry and fluctuations in random permutations \cite{rz}, 
biological systems \cite{kwa}, stock markets \cite{pl}, atmospheric sciences 
\cite{santh}, complex networks \cite{da} etc (see also \cite{co1}).
The stationary random matrix ensembles are basis-invariant ensembles,  
characterized by similar and independent distribution of almost all elements 
\cite{me}. This restricts their applicability only to the generators with 
wave-functions extended in the entire system or with a coherent scattering 
of waves. However the matrix elements distribution can significantly be 
affected by various system conditions e.g missing interactions among some 
of the sub-systems, a variation in their degree or nature, 
symmetry and boundary conditions, dimensionality, disorder etc 
\cite{psand}.  These condition may result in different strengths of the  
elements, correlations between them and localized waves; the corresponding 
ensembles are then basis-non-invariant.

The presence of local interactions and 
wave-localization phenomena is quite generic to a wide range of complex systems. 
The statistical analysis of their physical properties requires  therefore a search 
for new mathematical tools. The present study is an attempt in this 
direction. The basic idea here is to take into account the inaccuracy 
in the matrix representation of an operator of a 
complex system. The fluctuation of accuracy  
with changing system conditions results in a change of distribution 
parameters of various matrix elements. This leads to a 
seemingly multi-parametric diffusion of the ensemble density. 
However as shown here, the diffusion is essentially governed by a single 
parameter only. The information can then be  utilized to study the fluctuations 
of the physical properties due to varying system conditions and express the 
results in a common mathematical form for a wide range of complex systems. 

%	The results obtained by the accuracy based approach are analogous to 
%the maximum entropy models of complex systems \cite{ba, me, ps, psco}. 
%Recently the approach was used to derive the common mathematical formulation 
%for statistical measures of the complex systems  modeled by a multi-parametric 
%Gaussian probability density  and independent/ correlated matrix elements 
%\cite{ps, psco}. These results indicated the possibility of a classification of 
%the complex systems into various universality classes (based on the behavior of 
%their statistical measures and characterized by the complexity parameter). 
%However the maximum entropy approach is based on  the description 
%of the ensemble density, based on the extent of available information about 
%system conditions. It is relevant to seek a physical understanding of   
%the  underlying connection among complex systems which gives rise to various 
%common features in their statistical behavior. It is also desirable to explore the 
%possibility of generalization of the results to cases with non-polynomial 
%densitis  of the matrix elements. 

	The statistical behavior of complex systems and the possibility of a 
common mathematical formulation was recently studied by considering their 
maximum entropy models \cite{ps,psco}.  The latter are based on the  
formulation 
of the ensemble density by maximizing the information entropy under 
constraints imposed on the system \cite{ba}. The ensemble density is then 
utilized to extract the distribution of eigenvalues and eigenfunctions and 
desired physical information. The maximum entropy approach indicated the 
possibility of a classification of the complex systems into various 
universality classes (based on the behavior of their statistical measures 
and characterized by the complexity parameter \cite{ps,psco}). These 
results were also verified numerically for certain cases 
\cite{psand, pswf, pssymp}. However the complexity parameter formulation  
within this approach gives rise to some queries which required a more 
intuitive physical reasoning for their resolution.  
This motivates us to consider the accuracy based approach 
which not only resolves the queries but also helps  
in generalization of the single parametric formulation to a wider 
range of complex systems (those with system conditions subjecting 
 matrix elements to a potential with a single 
minima only).

	The paper is organized as follows. Section II describes the diffusive  
dynamics for the matrix elements of a Hermitian operator subjected to  
an external potential of type ${\rm e}^{-V(H)}$ as well as random noise 
originated in complexity of the system. (We have considered here the 
real-symmetric case only however the results are valid for 
complex Hermitian and real-quaternion cases too). 
The comparison of this approach 
with maximum entropy approach is discussed in section III. 
This is followed by section IV describing the derivation 
of the statistical measures of the eigenvalues and eigenfunctions using 
standard perturbation theory. Note, for generic potentials, the 
derivation of the measures by a direct integration of the evolution 
equation (the method used  for Gaussian cases in \cite{pswf}) 
is technically difficult. 
We conclude in section V with a summary of our main results.

\section{Accuracy Driven Diffusion of Matrix Elements } 

%	The idea of this approach originates from the Dyson's Brownian 
% motion model for the evolution of a system subjected to 
% random perturbation due to varying perturbation strength. 

	Consider, as an example, a Hermitian operator $H$ of
a complex system with time-reversal symmetry and integer angular momentum.
It is possible to choose a generic basis, say $|\phi_k \rangle$
($k=1 \rightarrow N$),  preserving time-reversal symmetry
for the matrix representation of $H$; the matrix turns out to be
real-symmetric in this basis with its elements
$H_{kl}=\langle \phi_k | H \phi_l \rangle$. For notational simplification,
let us denote them by $H_{\mu}$ where $\mu\equiv \{kl;s\}$ is a single index
which can take a value from $1\rightarrow { M}$
(${ M}=N(\beta N-\beta+2)/2$ the number of independent matrix elements).
Here $\beta$ is the number of commponents of $H_{\mu}$; 
thus $\beta=1$ for the real-symmetric case.

	Due to presence of complicated interactions in the system, it 
is technically difficult to evaluate some/all elements of the operator matrix 
in a generic basis. Consequently the matrix elements can be determined only 
within a certain degree of 
accuracy which, being sensitive to local system conditions, varies 
from element to element. The accuracy fluctuates rapidly as system 
conditions change, with different "time-scale" of fluctuations for 
each matrix element. The variation of an element $H_{\mu}$ with 
changing system conditions can therefore be mimicked by a  
particle undergoing Brownian dynamics due to rapidly fluctuating forces 
in addition to an external force (due to existing system conditions). 
The matrix elements of a physical system also have a natural tendency 
to oppose the cause for their change. The dynamics is therefore subjected 
to a local frictional force too.

  Consider the "particle" $H_{\mu}$ in equilibrium under external 
force $V(H_{\mu})$ due to existing system conditions. 
The equation of motion for $H_{\mu}$ due to changing system conditions 
can be written as 
\begin{eqnarray}
{{\rm d}^2 H_{\mu} \over {\rm d}t_{\mu}^2}   = 
-f {{\rm d} H_{\mu} \over {\rm d}t_{\mu}} + V(H_{\mu}) +A(t_{\mu})
\label{e1}
\end{eqnarray}

where $f$ is the friction coefficient and $A(t_{\mu})$ is a rapidly fluctuating 
force in "time" $t_{\mu}$ (a pseudo time only, a measure of the scale for 
accuracy fluctuations) with following usual properties:

\begin{eqnarray}
\langle A(t_{\mu 1}) A(t_{\mu 2})...A(t_{\mu (2n+1)}) \rangle &=& 0, \label{e2} \\
\langle A(t_{\mu 1}) A(t_{\mu 2})...A(t_{\mu (2n)}) \rangle &=& 
\sum_{pairs} \langle A(t_{\mu i}) A(t_{\mu j})\rangle 
\langle A(t_{\mu k}) A(t_{\mu l}) \rangle...., \label{e3} \\
\langle A(t_{\mu i}) A(t_{\mu j})\rangle &=& 
(2/ f ) \delta(t_{\mu i}-t_{\mu j}). 
\label{e4}
\end{eqnarray}
where $\langle . \rangle$ refers to ensemble average, 
$t_{\mu j}$ refers to $j^{th}$ step in "time"-scale $t_{\mu}$ and the 
summation in eq.(\ref{e3}) extends over all distinct ways in 
which the $2n$ indices can be divided into $n$ pairs. Further,  
for a clear exposition of the ideas, we consider here the potential 
$V(H_{\mu})$ as a function of $H_{\mu}$ with a single minima.
  
. 

The Langevin equation can now be integrated: let $H_{\mu}$ be the position 
of the particle at "time" $t_{\mu}$ which changes to position 
$H_{\mu} +\delta H_{\mu}$ at a later "time" $t_{\mu} +\delta t_{\mu} $ 
(here $t_{\mu}$ chosen to be long enough for the effects of initial velocity 
to become negligible). Due to presence of 
rapidly fluctuating forces, the variation $\delta H_{\mu}$ 
 in position of the particle will behave like a 
random variables. Using eqs.(\ref{e1}-\ref{e4}) and keeping 
terms only of first order in $\delta t_{\mu}$, one gets

\begin{eqnarray}
f\langle \delta H_{\mu} \rangle = - V(H_\mu) \delta t_{\mu},  
\qquad \qquad 
f\langle (\delta H_{\mu})^2 \rangle = (g_{\mu}/\beta) \delta t_{\mu},  
\label{e5}
\end{eqnarray}
with $g_{\mu} \equiv g_{kl} = 1+ \delta_{kl}$.   
  Due to random variations in particle position with changing system 
  conditions, it is 
  appropriate to consider a "time"-dependent probability density 
  $\rho_{\mu} (H_{\mu}, t_{\mu})$ that 
  the particle will be at the position $H_{\mu}$ at "time" $t_{\mu} $. Assuming 
  a Markovian process (that is the independence of future evolution 
  from past states, dependence only on present state), one can write 

\begin{eqnarray}
\rho_{\mu}(H_{\mu}; t_{\mu} +\delta t_{\mu} ) = \int 
\rho_{\mu}(H_{\mu}-\delta H_{\mu} ; t_{\mu}) 
\rho_{cond}(H_{\mu}-\delta H_{\mu}; \delta H_{\mu} ; \delta t_{\mu}) 
\; \; {\rm d}\delta H_{\mu}
\label{e6}
\end{eqnarray}
where $\rho_{cond}$ is the conditional probability that the position of 
the particle changes from $H_{\mu}-\delta H_{\mu}$ to $H_{\mu}$ in a 
time interval 
$\delta t_{\mu} $. Expanding both sides of eq.(\ref{e6}) in a power series 
of $\delta H_{\mu}$ and $\delta t_{\mu}$ and subsequently using eq.(\ref{e5}) 
we get (in limit $\delta t \rightarrow 0$)

\begin{eqnarray}
f {\partial \rho_{\mu} \over \partial t_{\mu}} =  
{\partial \over \partial H_{\mu}} \left[ {g_{\mu}\over 2\beta} 
{\partial \over \partial H_{\mu}} + V(H_{\mu}) \right] \rho_{\mu} 
\label{e7}
\end{eqnarray}

Equation(\ref{e7}) describes the evolution of $H_{\mu}$ with respect 
to "time"-scale $t_{\mu}$ which in turn depends on the "time"-scale 
for accuracy-fluctuations (and therefore system conditions) 
surrounding $H_{\mu}$. For systems where the coupling of  any two basis 
states through the generator $H$ is independent of coupling between other 
states (i.e all matrix elements are independent of each other), the 
fluctuations in accuracy of each matrix element are independent too. 
Each element can therefore be assumed to be subjected to a random  
force fluctuating at a time-scale independent of others, (that is, all $t_{\mu}$ 
independent of each other). This gives us $M$ equations, 
of type (\ref{e7}), for the independent evolutions of M elements $H_{\mu}$.

The joint probability distribution  $\rho(\{H_{\mu}\};\{ t_{\mu} \})$ of all 
matrix elements can now be defined as

\begin{eqnarray}
\rho(\{H_{\mu}\};\{t_{\mu} \}) = \prod_{\mu} \rho_{\mu} (H_{\mu} ;t_{\mu})
\label{e8}
\end{eqnarray}

which along with eq.(\ref{e7}) leads to the equation for  multi-parametric  
evolution of $\rho$: 

\begin{eqnarray}
f \sum_{\mu} {\partial \rho \over \partial t_{\mu}} =  \sum_{\mu} 
{\partial \over \partial H_{\mu}}\left[ {g_{\mu}\over 2 \beta} 
{\partial \over \partial H_{\mu}} + V(H_{\mu})  \right] \rho
\label{e9}
\end{eqnarray}

       For a system undergoing evolution as a whole unit, it is natural to 
seek a common scale, say $\tau$, at which all its constituents i.e matrix 
elements vary simultaneously. Let us therefore consider the evolution of 
$\rho$ with respect to $\tau$.  Assuming again a Markovian process, 
we have 

\begin{eqnarray}
\rho(\{H_{\mu}\}; \tau +\delta \tau ) = 
\int \rho(\{H_{\mu}\}-\delta \{H_{\mu}\} ; \tau) 
\rho_{cond}(\{H_{\mu}\}-\delta \{H_{\mu}\}; \delta \{H_{\mu}\} ; \delta \tau) 
\; {\rm D}\delta H
\label{e10}
\end{eqnarray}
where ${\rm D}\delta H \equiv \prod_{\mu}{\rm d}\delta H_{\mu}$
Expanding both sides of eq.(\ref{e10}) in a power series of 
$\delta H_{\mu}$ and $\delta \tau $, 
we get (in limit $\delta t \rightarrow 0$)

\begin{eqnarray}
 {\partial \rho \over \partial \tau } \delta \tau = 
 \sum_{\mu} {\partial \over \partial H_{\mu}}
\left[ {\partial \over \partial H_{\mu}} 
{\langle (\delta H_{\mu})^2 \rangle \over 2}
- \langle \delta H_{\mu} \rangle \right] \rho
\label{e11}
\end{eqnarray}
As both eq.(\ref{e11}) and eq.(\ref{e9}) describe the evolution of 
the probability density of $H$, they should be analogous. 
A comparison of the equations then gives the conditions
\begin{eqnarray}
{\partial \rho \over \partial \tau } \delta \tau &=& 
 \sum_{\mu} {\partial \rho \over \partial t_{\mu} } \delta t_{\mu} 
\label{e11+} 
\end{eqnarray} 
and 
\begin{eqnarray}
f\langle \delta H_{\mu} \rangle &=& - V(H_{\mu}) \delta \tau,  \nonumber \\
f\langle (\delta H_{\mu})^2 \rangle &=& (g_{\mu}/\beta) \delta \tau,  
\label{e12}
\end{eqnarray}
The two conditions imply
\begin{eqnarray}
\delta \tau &=& \delta t_1 = \delta t_2 = ..= \delta t_{\tilde M}.
\label{e11++}
\end{eqnarray} 
The above is satisfied if 
$\tau$ is defined as $\tau = {\sum_{\mu=1}^N a_{\mu} t_{\mu} \over 
\sum_{\mu} a_{\mu}}$ with $a_{\mu}$ as arbitrary constants. However 
physical reasoning (based on no preference by random forces to any 
particular component of the system) suggests us to choose $a_j$ equal.

The solution of eq.(\ref{e11}) for arbitrary initial condition, say $H_0$ at 
$\tau=\tau_0$ can be given as 
\begin{eqnarray}
\rho(H,\tau|H_0,\tau_0) \propto {\rm exp}[-(\alpha/2 \beta) {\rm Tr}(H- \eta H_0)^2]
\label{e13}
\end{eqnarray}
with
$\alpha= (1-\eta^2)^{-1}$ and $\eta={\rm e}^{- (\tau-\tau_0)/f}$. 
The probability density of $H$ can now be extracted by integrating over an 
ensemble of initial conditions. Although eq.(\ref{e11}) and eq.(\ref{e13}) 
are derived for the case $\beta=1$, it is easy to show, following essentially the 
same steps, their validity  for the complex Hermitian case $\beta=2$ 
and real-quaternion case $\beta=4$. 
 
%Eq.(\ref{e11}) describes a single parametric evolution of the matrix elemnts 
%density $\rho(H)$. 

Note the accuracy scales $\tau_{\mu}$ depend on local 
system conditions which can vary from system to system. However as 
eq.(\ref{e11}) and eq.(\ref{e13}) indicate, $\rho(H)$ is insensitive to 
the details of the local system conditions; it depends only on their 
average behavior described by $\tau$ besides global constraints 
e.g. $V(H)$ and symmetry conditions; (Note $V(H)$ has no explicit 
dependence on $\tau_{\mu}$). 
Thus, analogous to their maximum 
entropy models, the accuracy based approach indicates   
a single parametric dependence of the density $\rho(H)$ for simple 
harmonic confinement $V(H)=H$. It further 
generalizes the formulation to the systems with conditions giving 
rise to a generic single-well (single minima) potential. 
The approach can in principle be extended to the multi-well potentials 
 too however it requires a modification of the technical details.  
We intend to pursue these cases in near future.

It is important to note that the form of eq.(\ref{e11}) for case $V(H)=H$ is 
analogous to Dyson Brownian model \cite{dy, me}. The latter deals with the 
case of a stationary ensemble subjected to a random perturbation.
However the Brownian dynamics of matrix elements in accuracy model 
is different from Dyson's case; there are two main differences:

(1) In Dyson's model, the randomness caused due to a perturbation
 is same for almost all  matrix elements. In accuracy model, the origin of
randomness is the lack of accuracy which is sensitive to local
conditions. Different matrix elements therefore may be subjected to 
different randomness.
                                                                                  
 (2) In Dyson's model, the evolution occurs due to a variation in the 
     perturbation strength and is single parametric. In accuracy model,
     the evolution is brought by the fluctuating accuracy due to varying
     system conditions. As a consequence, we need to consider a
     multi-parametric evolution of probability density (unlike single
     parametric evolution in Dyson's case). However, as eq.(\ref{e11}) 
     indicates, the multi-parametric evolution can be reduced to a 
     single parametric evolution.

%here the differences between accuracy based approach leading to Brownian 
%dynamics of matrix elements and Dyson's Brownian motion model 
%However there are two main differences:

% However the latter can be fixed by the conditions on moments 
%following eq.(\ref{e5}) 
% which implies that 

%A simultaneous validity of eq.(\ref{e12}) and eq.(\ref{e5}) (for all $\mu$) 
%requires $\delta \tau = \delta t_{\mu}$ (for all $\mu$) and therefore $a_{\mu}=1$ 
%(as $\delta \tau =\sum_{\mu} a_{\mu} \delta t_{\mu} /(\sum_{mu} a_{\mu})$.  

%	For $F(H_{\mu})=H_{\mu}$, eq. (\ref{e11}) is same as eq.(\ref.{c9}) 
%with $\tau$ replaced by $Y$. Thus the consideration of accuracy as rapidly 
%fluctuating forces acting on the operator of a complex system leads to same single 
%parametric formulation as obtained by their Gaussian ensemble modelling based on 
%maximun entropy hypothesis. Note for $F(H_{\mu})=H_{\mu}$, 
%the solution of the eq.(\ref{e11}) is a Gaussian. 

%eq.(\refe5})

%Starting from an arbitrary initial probability density $P_j$ at time 
%$t_j=t_0$, a unique solution of eq.(\ref{e7}) will exist for all 
%$t_j \ge t_0$. 

%$x_1, x_2, .., x_n$ be the positions of the particle at time $t$ which 
%change to positions $x_1+\delta x_1$, $x_2 +\delta x_2,..,\delta x_n$ 
%at a later time $t+\delta t$ (here $t$ chosen to be long enough for the 
%effects of initial velocity to become negligible). Due to presence of 
%rapidly fluctuating forces, the variations $\delta H_{\mu}$ 
%($j=1 \rightarrow N$) in positions of the particles will behave like 
%random variables. Using eqs.(\ref{e1}-\ref{e4}) and keeping 

\section{Comparison of accuracy based approach and maximum entropy approach}

%The maximum entropy approach is based on the representation of a complex system 
%by an ensemble of matrices. Here the probability density of the matrix elements 
%is formulated by exploiting some prior information about the system.  For example, 
%for a system with known average behavior of the  matrix elements and their 
% variances, the ensemble density can be given as  

% exloits the Gaussian or polynomial form of the potential.
% The common mathematical formulation using maximum entropy approach is 
% discussed in detail 

% This section compares the results obtained in previous section to 
%those derived by maximum entropy approach in \cite{ps, psco}. To simplify, 
%the comparison is restricted to multi-parametrix Gaussian ensembles only. 

The objective of this section is to indicate the analogy of the 
results obtained by the accuracy model and maximum entropy 
models of complex systems notwithstanding their 
seemingly different origins. For a clear comparison, we briefly review 
the maximum entropy approach. 
This approach is based on the representation of a 
complex system by an ensemble of matrices; here the probability density 
of the matrix elements is formulated by maximizing the information entropy 
under known system-constraints (see \cite{ba} for details).  
However the density in accuracy based approach is 
obtained as a non-stationary state of a diffusion process. 
This can further be clarified by an example.
The accuracy model leads to a Gaussian density $\rho(H)$ 
if $V(H)=H$ and the initial density is Gaussian too 
(see eq.(\ref{e13}). 
However the maximum entropy theory leads to a Gaussian density 
if the available information about matrix elements is 
limited to their average behavior and variances only:  
\begin{eqnarray}
 \rho (H,\nu,b)= \prod_{\mu} \rho_{\mu}(H_{\mu}, v_{\mu}, b_{\mu}) = 
 C{\rm exp}[{-\sum_{\mu} (1/2 v_{\mu}) (H_{\mu}-b_{\mu})^2 }]
\label{c3}
\end{eqnarray}
with $C$ as a normalization constant, $v,b$ as the 
matrices of variances $v_{\mu}$ and mean $b_{\mu}$, 
respectively, and the symbol $\sum_{\mu}$ implying a summation 
over independent matrix elements only.

The emergence of single parametric formulation in 
maximum entropy approach can briefly be explained as follows. 
The Gaussian nature of $\rho_{\mu}$ (see eq.(\ref{c3}))  
leads to  a relation among its derivatives with respect to 
$H_{\mu}, v_{\mu}, b_{\mu}$:

\begin{eqnarray}
T_{\mu} \rho = L_{\mu} \rho 
\label{c4}
\end{eqnarray}
where 
\begin{eqnarray}
T_{\mu} &=& \left[(2/\tilde g_{\mu}) 
x_{\mu}{\partial \rho\over\partial v_{\mu}} - 
  b_{\mu} {\partial \rho\over\partial b_{\mu}}\right]   \label{c5} \\
L_{\mu} &=&  {\partial \over \partial H_{\mu}}
\left[ {g_{\mu}\over 2\beta} {\partial \over \partial H_{\mu}} +
 H_{\mu} \right] \rho
\label{c6}
\end{eqnarray}
with $x_{\mu} \equiv 1- {\tilde g_{\mu}} v_{\mu}$ with 
${\tilde g_{\mu}} \equiv {\tilde g_{kl}} =2-\delta_{kl}$ and 
$g_{\mu}$ same as in eq.(\ref{e5}). 
A particular combination of the parametric derivatives $T=\sum_{\mu} T_{\mu}$ 
leads to a diffusion equation $T\rho= L\rho$ (with $L=\sum_{\mu} L_{\mu}$).  
The single parametric formulation of the diffusion then follows by showing 
$T = {\partial \over \partial Y}$ with $Y$ as the complexity parameter 
\cite{ps,psco}.

% The above equation implies a parametric diffusion of $H_{\mu}$ 

% As mentioned above, the case  $V(H_{\mu})=H_{\mu}$ in accuracy 
% model leads to a Gaussian density for $\rho(H_{\mu})$. 

Eq.(\ref{c4}) describes the evolution of $\rho$ when all other matrix 
elements except $H_{\mu}$ is held fixed. It is therefore equivalent 
to eq.(\ref{e7}) with $V(H_{\mu})=H_{\mu}$ and $\rho_{\mu}$ replaced 
by $\rho$ (following eq.(\ref{e8}). This implies     
\begin{eqnarray}
f{\partial \rho \over \partial t_{\mu}} =
(2/\tilde g_{\mu})
x_{\mu}{\partial \rho\over\partial v_{\mu}} -
b_{\mu} {\partial \rho\over\partial b_{\mu}}
\label{c7}
\end{eqnarray}
The scale $t_{\mu}$ can then be expressed in term of 
the distribution parameters:
\begin{eqnarray}
v_{\mu} &=&  
(1-{\rm e}^{2 t_{\mu}/f})/{\tilde g_{\mu}} + c_{1 \mu} , \nonumber \\
b_{\mu} &=& {\rm e}^{t_{\mu}/f} + c_{2 \mu} . 
\label{c8}
\end{eqnarray}
with $c_{1 \mu}$ and $c_{2 \mu}$ as 
constants specific to each $v_{\mu}$ and $b_{\mu}$ respectively.
The above indicates the equivalence of $\tau=\sum_{\mu} t_{\mu}$ 
(the average scale for accuracy fluctuation) to complexity parameter 
in maximum entropy model (an average distribution parameter of the ensemble).  
This further implies that the  confinement by a 
simple harmonic force in the accuracy  model is equivalent to the maximum  
entropy modeling of a system with known averages and variances of the 
matrix elements. 
Similarly a general confining potential $V(H)$ in accuracy model can be 
shown to be  equivalent to a maximum entropy ensemble derived under 
constraints $\langle U(H) \rangle$ =constant where $U(H)=\int V(H) {\rm}dH$.

%This is consistent with our description of $\tau=\sum_{\mu} t_{\mu}$ 
%as the average scale for accuracy fluctuation in eq.(\ref{e11}). 

%The accuracy approach leads to the definition 
%of $\tau \equiv Y$ as the average scale for the accuracy fluctuations. 

% single parameter, known as complexity parameter, which turns out to 
% be average distribution parameter of the ensemble  measured in units 
% of mean level spacing.  

	The equivalence of the accuracy approach to maximum entropy 
approach can  be used to clarify some of the points related to latter.    
For example, in maximum entropy approach, a particular combination $T\rho$ of 
the parametric derivatives leads to Brownian type diffusion; the reason 
to consider such a combination are not so obvious. However the accuracy 
approach clearly explains the reason: the combination is 
required to study the evolution of the system as a whole unit.  
%
%it leads to a simultaneous diffusion of the matrix elements  
%due to random forces arising from  uncertainties.
%
Further, in the maximum entropy approach, the multi-parametric 
diffusion governed by the parameters $Y_j$, $J=1 2,...N$ is 
reduced to a single parametric formulation by showing that all 
$Y$'s except $Y_1$ are constants of evolution. 
However, in accuracy based approach, 
the single parameter existence follows from necessity of the 
simultaneity of the dynamics of various matrix elements. As  
both approaches represent the same dynamics,  
this reconfirms the lack of any role 
played by the parameters $Y_2..,Y_M$ in the diffusion of matrix elements.

%Further	
%For $F(H_{\mu})=H_{\mu}$, the solution of the eq.(\ref{e11}) is a 
%Gaussian which can also be written in 
%the form of eq.(\ref{c3}) using eq.(\ref{e14}). The  confinement by a  
%simple harmonic force in the accuracy  model 
%therefore corresponds to the maximum  
%entropy ensemble modeling of a complex system with 
% first two moments of its matrix elements subjected to known constraints. 
%A general $V(H)$ therefore corresponds to a maximum entropy model 
%derived by exploiting known constraints on any function of its matrix 
%elements. 

\section{Diffusion of Eigenvalues and Eigenfunctions}

The eigenvalue equation of a $N\times N$ Hermitian matrix $H$ is 
given by $H U = U^{\dagger} E$ with $E$ as the $N\times N$ diagonal matrix of 
eigenvalues,  $E_{mn}= e_n  \delta_{mn}$ and $U$ as the $N\times N$ 
eigenvector matrix, unitary in nature: $U^T . U=1$ \cite{me}. As 
described in \cite{pswf}, the statistics of the eigenvalues and/or 
eigenfunctions of $H$ can be obtained from eq.(\ref{c5}) by integrating 
over the eigenfunctions, eigenvalues respectively; the results in \cite{pswf} 
however are valid only for $V(H)$ as a simple harmonic force. Here we apply 
 2nd order standard perturbation theory \cite{landau} to derive results for 
a more general form of $V(H)$; here again $H$ is taken to be a real-symmetric 
matrix for simplification.

\subsection{Eigenvalue Statistics}

A small change $\delta \tau$ in the parameter $\tau$ changes $\rho(H)$ and its 
eigenvalue statistics. By considering matrix $H+\delta H$ in the diagonal 
representation of matrix $H$,
the change $\delta e_n$ in the eigenvalues can be given as
\begin{eqnarray}
\delta e_n = \delta H_{nn} + 
\sum_{m\not=n} {|\delta H_{mn}|^2 \over e_n-e_m}+
o((\delta H_{mn})^3)
\label{f1}
\end{eqnarray}
where $H_{mn}=e_n \delta_{mn}$ at value $\tau$ of the parameter.
This further gives,
\begin{eqnarray}
f\langle{\delta e_n}\rangle  &=& \langle {\delta H_{nn}}\rangle +
\sum_{m=1,m\not=n}^{N} {\langle {|\delta H_{mn}|^2}\rangle \over e_n-e_m} \\
&=& \left[- V(e_n) +  
\sum_{m=1,m\not=n}^N {1 \over e_n-e_m}\right] \delta Y 
\label{f2}
\end{eqnarray}
Here eq.(\ref{f2}) has been obtained from eq.(\ref{f1}) by 
using eq.(\ref{e12}). 
Similarly, upto  first order of $\delta \tau$,

\begin{eqnarray}
f \langle{\delta e_n \delta e_m }\rangle =
\langle{\delta H_{nn} \delta H_{mm}}\rangle =
 (2/\beta)  \delta_{nm} \delta \tau
\label{f3}
\end{eqnarray}

%The eq.(7) can be obtained from eq.(18), by following standard routes
%of calculation of the moments from a Fokker-Planck equation; 
%see \cite{meta,vk} for details.

The information about moments of the eigenvalues $e_n$ can
now be used to obtain their evolution equation. The theory of Brownian
motion \cite{vk} informs us that the joint probability distribution
$P(\{e_n\})$ for the eigenvalues $e_n$ evolves with increasing
$\tau$ according to Fokker-Planck equation,

%{\partial P \over\partial Y} {\delta Y}  &=&
%\sum_{n=1}^N{\partial \over \partial \lambda_n}\left[{1\over 2}
%{\partial (\overline{\delta\lambda_n\delta\lambda_m}) 
% P\over \partial \lambda_n} +
%(-\overline{\delta\lambda_n})\; P \right] \nonumber \\

\begin{eqnarray}
\beta f{\partial P \over\partial \tau}   &=& L_E P \label{f3+} \\
L_E &=&  \sum_{n=1}^N
{\partial \over \partial e_n}\left[ {\partial \over \partial e_n} +
\sum_{m=1,m\not=n}^N {\beta \over e_m-e_n} + \beta V(e_n)  \right] P
\label{f4}
\end{eqnarray}
The above equation  describes the evolution of the eigenvalues of a complex 
system modeled by the ensemble $\rho(H)$ due to changing system conditions.

%Note it

%.

 	As in the case of maximum entropy approach \cite{ps,psand,ap}, 
the eigenvalue correlations for 
the case $V(e)=e$ can be obtained by using the analogy with Dyson's Brownian 
ensembles \cite{dy, me}. For a general $V(e)$, the correlations can be analyzed 
by mapping the eq.(\ref{f3+}) to Calogero-Sutherland  Hamiltonian \cite{ap}.  
This can be achieved by using the transformation $\Psi=P/|Q_N|^{1/2}$ in 
eq.(\ref{f3+}) reducing it in a form 
${\partial \Psi \over\partial \tau} = -\hat H \Psi$
with $Q_N=\prod_{m\not=n} (e_m-e_n)^{\beta} 
{\rm exp}[-\beta \sum_n U(e_n)]$ and $U(e)= \int {\rm d}e V(e)$.
The 'Hamiltonian' $\hat H$ turns out to be the Calogero-Sutherland 
Hamiltonian in one dimensions \cite{ap} 

\begin{eqnarray}
\hat H = -\sum_i  {\partial^2 \over \partial e_i^2} +
 \sum_{i,j; i<j} {\beta (2-\beta)\over (e_i-e_j)^2} + \beta \sum_i  V(e_i)
\label{f5}
\end{eqnarray}

Similar to the case $V(e)=e$ (see \cite{ap,ps} for details), the 
"state" $\psi$ or $P(\{e\},\tau| H_0)$ for 
a generic  $V(e)$ can be expressed as a
sum over the eigenvalues and eigenfunctions of $\hat H$. The  
integration of the sum over  the initial ensemble  $H_0$ would then lead to joint 
probability distribution $P(\{e\},\tau)$ and
thereby density correlations $R_n$ for unfolded spectrum (eigenvalues 
rescaled in units of local spectral density). Note the choice of initial 
eigenvalue distribution at $\tau_0$ depends on the global system constraints. 

	As eq.(\ref{f4}) and eq.(\ref{f5}) indicate, the confining 
potential $V(e)$ does not affect the short range level-correlations. 
The latter are governed only by the complexity parameter $\tau$ and 
underlying exact symmetry conditions.   
However the long range level-correlations are sensitive to two factors: 
(i)  the complexity parameter (i.e the average accuracy fluctuation scale 
$\tau$, or equivalently,  the average distribution parameter), 
(ii) the global system constraints i.e details of external force $F(H)$ and 
the symmetry conditions. 
Thus the systems subjected to {\it effectively} similar physical 
constraints will show analogous long-range correlations 
(after spectral-unfolding) if their complexity parameter are equal. 
Here the term "{\it effectively} similar physical constraints" 
implies the similar symmetry conditions as well as same mathematical form of 
the external potential although 
it may originate from different physical conditions. 
For example, a harmonic confinement of the matrix elements 
which also corresponds to their Gaussian distribution can be  
a physical characteristic of many systems related to different 
areas of physics.

\subsection{Eigenfunction Statistics}
	    
	The evolution equation for the probability density of various 
eigenfunction components can similarly be obtained. Here again we 
consider the case of a real symmetric operator for simplification. 
The eigenvector matrix $U \equiv O$ is then orthogonal: 
$O^T . O=1$ \cite{me}. Using standard 
perturbation theory for Hermitian operators, the second order change 
in $j^{\rm th}$ component $O_{jn}$ of an eigenfunction $O_{n}$ due to a 
small change $\delta \tau$ can be described as 
  
\begin{eqnarray}
\delta O_{jn} = \sum_{m\not=n} {|\delta H_{mn}| \over e_n-e_m} O_{jm} +
\sum_{m,m'\not=n}^{N} {\overline |\delta H_{mn}|  |\delta H_{m'n}|  
\over (e_n-e_m)(e_n-e_{m'} )} O_{jm} \nonumber \\
-\sum_{m\not=n}^{N} {\overline |\delta H_{mn}|  |\delta H_{nn}|  
\over (e_n-e_m)^2} O_{jm} 
- {1\over 2} O_{jn} \sum_{m\not=n}^{N} {\overline |\delta H_{mn}|^2  
\over (e_n-e_m)^2} 
\label{f5+}
\end{eqnarray}
As eq.(\ref{e12}) indicates, the matrix elements of $H$ are uncorrelated. 
Furthermore, at $\tau$, $H_{mn}= e_n \delta_{mn}$ (due to $H+\delta H$ 
being considered in diagonal representation of $H$) which gives, following 
from eq.(\ref{e13}),  
$\delta H_{mn}=-V(H_{mn}) \delta \tau = -V(e_n \delta_{mn}) \delta \tau$. 
Thus $\delta H_{mn}=V(0) \delta \tau =0$ for $m\not= n$ and $V(0)=0$.  
The ensemble averaged $O_{jn}$ 
then has a non zero contribution only from the last term of 
eq.(\ref{f5+}) (see eq.(\ref{e5})):

\begin{eqnarray}
\langle{\delta O_{jn}}\rangle &=& 
= - {1\over 2}   
\sum_{m=1,m\not=n}^N {O_{jn}  \over (e_n-e_m)^2} \delta \tau 
\label{f6}
\end{eqnarray}
Note for cases where $V(H_{mn})$ is nonzero for $m\not=n$, the first 
term contributes too.  
Further for cases where $V(0) \not=0$ or matrix elements are correlated, 
the other terms may also contribute.

The $2^{nd}$ moment of the eigenvector components has a contribution 
only from the first term  in eq.(\ref{f5+}) (up to first order in $\delta \tau$)  
\begin{eqnarray}
\langle{\delta O_{jn} \delta O_{kn} } \rangle =
\sum_{m,m'\not=n}^{N} {\langle |\delta H_{mn}|  |\delta H_{m'n}| \rangle  
\over (e_n-e_m)(E_n-E_{m'} )} O_{jm} O_{km'}
=  \sum_{m=1,m\not=n}^N {O_{jm} O_{km} \over (e_n-e_m)^2} \delta \tau  
\label{f7}
\end{eqnarray}

As the moments for eigenfunction components depend on eigenvalues too, we can 
first write the diffusion equation for the joint probability density 
$P_{ef,ev} (e_1,e_2,..,e_n;Y)$ of all the components of an eigenfunction and 
all eigenvalues:

\begin{eqnarray}
{\partial P_{ef,ev} \over\partial \tau} &=& ( L_O + L_E) P_{ef,ev} 
\label{f8} 
\end{eqnarray}
where $L_O$ and $L_E$ refer to two parts of Fokker-Planck operator 
corresponding to eigenvalues and eigenfunction components. Here $L_E$ 
is given by eq.(\ref{f4}) and 

\begin{eqnarray}
L_O &=& \sum_{j} {\partial \over \partial O_{jn}}
\left[ {\partial \over \partial O_{jn}} \langle (\delta O_{jn})^2 \rangle
+ \langle \delta O_{jn} \rangle \right] 
\label{f9}
\end{eqnarray}
A substitution of the moments (eqs.(\ref{f6},\ref{f7})) in eq.(\ref{f9}) 
followed by an integration of eq.(\ref{f8}) over all eigenvalues except $e_n$  
will then lead to the evolution equation for joint probability density 
$P_n (O_n, e_n; Y)$; the equation turns out to be same as 
eq.(18) given in \cite{pswf} and can further be used to derive various 
correlation measures for an eigenfunction \cite{pswf}. 

 The eq.(\ref{f5+}) can also be used to derive the joint probability 
distribution of the components of different eigenfunctions; again, for 
the cases with $V(0)=0$,  the results, e.g. single parametric formulation 
in infinite size limit, turn out to be same as given in \cite{pswf}. 

%Note that eiegnfunction statistics remain unaffected by the presence of 
%external potential $F(H_{\mu})$

\section{conclusion}  

%The complexity parametric formulation of the statistical properties 
%suggests a deep-rooted universality hidden underneath the world of 
%complex systems. In this paper, we have presented a physically 
%motivated derivation of the complexity parameter. This is 
%achieved by treating the uncertainty associated with an operator of  
%a complex system as a random force leading to a diffusion of its matrix 
%elements. The derivation not only gives a better insight in the results 
%given by maximum entropy model approach, it generalizes them to a wider 
%range of systems. We find that the complexity parameter is the typical scale 
%for the accuracy fluctuation  of the matrix elements
%and is equivalent to the average distribution parameter of the 
%maximum entropy ensemble.
 
% essentially the same as the one implied by maximum entropy approach, 
% namely,  as the average distribution parameter of the ensemble. 

% Our analysis presented here is restricted to the conditions which subject 
% matrix elements to monotonic potentials. 
 
In this paper, we have studied the dynamics of the matrix elements of an 
Hermitian operator of a complex system subjected to a single-well 
potential. The dynamics is diffusive due to random forces 
originating from accuracy-fluctuations due to varying system conditions. 
The information is then applied to explore the statistical behavior 
of the eigenvalues and eigenfunctions. Our analysis suggests a possible 
classification of complex systems in an infinite range of universality 
classes characterized just by complexity parameter and the nature of global 
physical constraints. The constraints e.g. unitary/ antiunitray symmetries, 
%
%a bounded trace of the operator, fixed eigenvalue density, fixed moments etc.   
%
and confining potential on matrix elements 
seem to divide complex systems in various universality classes. Each such
class can further be divided into many sub-classes 
characterized by their complexity parameter. 
Note the "constraint" universality class of a system 
refers to the broad nature of its complexity (the finer details seem 
to be irrelevant). However its sub-universality class depends on the 
degree of complexity only (measured by complexity parameter). 
 This can be explained by following examples.      
The standard Gaussian orthogonal ensemble (GOE), power law ensemble 
of real matrices, and, time-reversal Anderson ensemble belong to same 
"constraints" universality class in the above classification \cite{pswf} 
although  their complexity parameters, in general, are not equal  
(approaching infinity for GOE and finite in the other two cases) .  
However for the system parameters leading to same finite value of 
the complexity parameter, the Anderson ensemble and power law  
ensemble show same statistics \cite{pswf}.

The accuracy approach described here is applicable, in its present form,  
 only to the cases with independent matrix elements subjected to a 
single-well potentials. The frequent occurrence of correlated elements or 
multi-well potentials among complex systems makes their analysis 
desirable too. A generalization to these cases 
requires a more involved technical analysis. However our intuition 
suggests the possibility of a similar classification for these cases too. 
For example, for the multi-well 
potentials, the accuracy scales and their fluctuations are sensitive to 
local system details and can therefore vary from one branch to another. 
This would lead to a variation of diffusion scales (the average 
accuracy scale or complexity parameter) in different branches. 
Thus the statistical properties within a single branch would belong 
to a universality class characterized by the local complexity parameter.  
However the universality classes in different branches need not be analogous. 
The above suggestion seems to be in accord with already known results 
for invariant ensembles with multi-well potentials \cite{ey}. This encourages 
us to pursue a detailed analysis and extension to non-invariant ensembles of 
such cases in near future.

For the correlated cases, the accuracy-scales for various 
elements are no longer independent. However, a recent study of 
the maximum entropy models of a few correlated cases indicates 
the existence of the universality classes among them too\cite{psco}. 
It is desirable to explore the 
possibility of its generalization to a wider range of such cases.

%	The Brownian motion approach described here is applicable only for 
% the cases with independent matrix elements. The generalization to cases 
% with correlated matrix elements would require introduction of the 
% time-scales  which govern the correlated dynamics. The single 
% parametric formulation of correlated systems has already been shown 
% within maximum entropy modeling however it would be useful to 
% study it through Brownian motion approach too.  

%	The approach descibed in section II is applicable only for 
% the cases with independent matrix elements. The generalization to cases 
% with correlated matrix elements would require introduction of the 
% time-scales  which govern the correlated dynamics. As the single 
% parametric formulation of correlated systems has already been shown 
% within maximum entropy modeling, it would be useful to 
% study it through Brownian motion approach.  

% This further suggests a deep-rooted universality hidden underneath the 
% world of complex systems.

%We have also discussed here the 
%complexity parameteric formulation of eigenvalue and eigenfunction 
%statistics using 2nd order perturbation theory  

%the same results for simple harmonic forces as the   

% We find thatThe consideration of accuracy associated 

\end{document}